\documentclass[Conference,a4paper]{IEEEtran}
\IEEEoverridecommandlockouts
\usepackage{siunitx}
\usepackage{color}
\usepackage{graphicx}
\usepackage{amsmath}
\usepackage{amssymb}
\usepackage{algorithm}
\usepackage{algorithmic}
\usepackage{amsmath}
\usepackage{multirow}
\usepackage{booktabs}
\usepackage{array}
\usepackage{amsthm}
\usepackage{stfloats}
\usepackage{caption}
\usepackage{subfigure}
\usepackage{bm}
\usepackage{epstopdf}
\usepackage{setspace}
\usepackage{gensymb}
\usepackage{url}
\usepackage{verbatim}   
\usepackage{balance}

\usepackage{cleveref} 
\usepackage[a4paper, left=1.3cm, right=1.3cm, top=1.81cm, bottom=4.43cm]{geometry}

\newcommand{\be}{\begin{equation}}
        \newcommand{\ee}{\end{equation}}
\newcommand{\bea}{\begin{eqnarray}}
        \newcommand{\eea}{\end{eqnarray}}
\newcommand{\ba}{\begin{array}}
        \newcommand{\ea}{\end{array}}

\allowdisplaybreaks[4]

\title{Impact of Insufficient CP on Sensing Performance in OFDM-ISAC Systems}
\author{ 
    \IEEEauthorblockN{Peishi Li$^{\dag}$, Rang Liu$^{\ddag}$, Qian Liu$^{\dag}$, and Ming Li$^{\dag}$\\}
	\IEEEauthorblockA{$^{\dag}$ Dalian University of Technology, Dalian, Liaoning 116024, China \\ E-mail: \texttt{lipeishi@mail.dlut.edu.cn, \{mli, qianliu\}@dlut.edu.cn}} \\
	\IEEEauthorblockA{$^{\ddag}$ University of California, Irvine, CA 92697, USA \\ 
		E-mail: \texttt{rangl2@uci.edu}} \vspace{-2mm}}

\begin{document}

\maketitle
\thispagestyle{empty}

\begin{abstract}
    Orthogonal frequency-division multiplexing (OFDM) is widely considered a leading waveform candidate for integrated sensing and communication (ISAC) in 6G networks. However, the cyclic prefix (CP) used to mitigate multipath effects in communication systems also limits the maximum sensing range. Target echoes arriving beyond the CP length cause inter-symbol interference (ISI) and inter-carrier interference (ICI), which degrade the mainlobe level and raise sidelobe levels in the range-Doppler map (RDM). This paper presents a unified analytical framework to characterize the ISI and ICI caused by an insufficient CP length in multi-target scenarios. For the first time, we derive closed-form expressions for the second-order moments of the RDM under both matched filtering (MF) and reciprocal filtering (RF) processing with insufficient CP length. These expressions quantify the effects of CP length, symbol constellation, and inter-target interference (ITI) on the mainlobe and sidelobe levels. Based on these results, we further derive explicit formulas for the peak sidelobe level ratio (PSLR) and integrated sidelobe level ratio (ISLR) of the RDM, revealing a fundamental trade-off between noise amplification in RF and ITI in MF. Numerical results validate our theoretical derivations and illustrate the critical impact of insufficient CP length on sensing performance in OFDM-ISAC systems.
\end{abstract}


\vspace{-3mm}
\section{Introduction}
Integrated sensing and communication (ISAC) has emerged as a key technology for sixth-generation (6G) wireless networks, enabling the simultaneous sensing of the environment and data transmission using shared hardware and spectrum resources \cite{LiuFan_JSAC_2022}. Among candidate waveforms, orthogonal frequency-division multiplexing (OFDM) stands out for its high spectral efficiency and robustness to frequency-selective fading, and has been widely adopted in existing standards. For radar sensing, OFDM enables effective decoupling of delay and Doppler estimations and exhibits a thumbtack-like ambiguity function \cite{Sturm_Proc_2011}, making it a leading candidate waveform for 6G ISAC systems.

In typical OFDM-ISAC systems, range-Doppler processing is performed using one of two primary filtering techniques: matched filtering (MF) or reciprocal filtering (RF). Several recent studies have evaluated the sensing performance of MF and RF through theoretical analysis and simulations, with a particular focus on key sidelobe metrics such as the peak sidelobe level ratio (PSLR) and integrated sidelobe level ratio (ISLR) \cite{Benmeziane_EuRAD_2022}-\cite{Keskin_TWC_2025}. A common assumption in these studies is that all target round-trip delays remain within the cyclic prefix (CP) duration, thereby ensuring that the received echoes are free from inter-symbol interference (ISI) and inter-carrier interference (ICI). However, this requirement imposes a stringent constraint on the maximum sensing range of the system \cite{Wu_JSAC_2022}. For example, with a subcarrier spacing of 120 kHz and a normal CP length as specified in the 5G NR standard \cite{standard_5GNR}, the interference-free range is limited to approximately 87.9 m, which is often insufficient for many practical sensing scenarios. It is therefore critical to investigate how the CP length affects the sensing performance in OFDM-ISAC systems.

To address scenarios where the CP is insufficient, some studies model the resulting distortion as additional noise. In particular, ISI/ICI arising from targets with round-trip delays exceeding the CP has been treated as an additive circularly symmetric complex Gaussian (CSCG) term, and closed-form expressions for the interference power have been derived under this assumption \cite{Wang_TVT_2025}, \cite{Xu_arxiv_2025}. However, those analyses focus exclusively on RF processing and do not consider how MF processing is impacted by an insufficient CP. As a result, a unified theoretical framework that comprehensively accounts for CP-induced ISI/ICI effects on both RF and MF processing remains to be established.

To fill this gap, this paper presents a unified analytical framework based on a generalized multi-target echo model that explicitly captures insufficient-CP-induced ISI and ICI when target delays exceed the CP duration. Within this framework, we derive exact closed-form expressions for the second-order moments of the range-Doppler map (RDM) for both RF and MF processing. To the best of our knowledge, this is the first derivation of such expressions for both filtering approaches. These results quantitatively reveal how the CP length, modulation constellation, and inter-target coupling jointly influence the mainlobe and sidelobe levels of the RDM. Building on these insights, we further derive closed-form expressions for the PSLR and ISLR, thereby illuminating the fundamental trade-off between interference suppression and noise amplification associated with different filtering choices and CP lengths. Finally, numerical results validate the accuracy of our theoretical predictions and demonstrate how an insufficient CP duration induces ISI/ICI that degrades the RDM's PSLR and ISLR for both RF and MF processing.

\vspace{-1mm}
\section{Signal Model and Echo Signal Processing} \label{sec:system_model}
\subsection{Transmit Signal Model}

\begin{figure}[!t]
    \centering
    \includegraphics[width = 3 in]{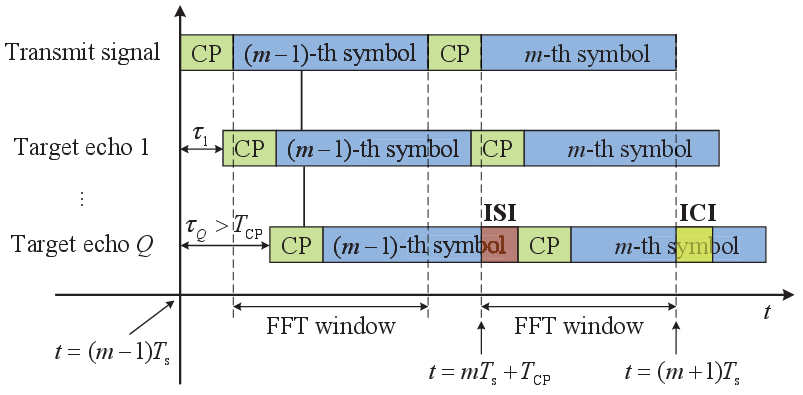}
    \caption{Illustration of transmit and echo signals. ISI and ICI occur when the CP length $T_{\text{CP}}$ is shorter than the maximum target delay $\tau_Q$.}
    \vspace{-4mm}
    \label{fig:echo_signal}
\end{figure}

Consider an OFDM frame with $M$ symbols and $N$ subcarriers, the baseband transmit signal including the CP can be expressed as \vspace{-1mm}
\begin{equation}
    x(t) = \sum_{m=0}^{M-1} \sum_{n=0}^{N-1} \frac{s_{n,m}}{\sqrt{N}} e^{\jmath 2 \pi n \Delta_f (t - T_{\text{CP}} - mT_{\text{s}})} g\Big(\frac{t - mT_{\text{s}}}{T_{\text{s}}}\Big), \vspace{-1mm}
\end{equation}
where $s_{n,m}$ denotes the data symbol on the $n$-th subcarrier of the $m$-th symbol, $\Delta_f$ denotes the subcarrier spacing, $T = 1/\Delta_f$ denotes the OFDM symbol duration,  $T_{\text{s}} = T + T_{\text{CP}}$ is the total symbol duration including the CP length $T_{\text{CP}}$, and $g(\cdot)$ denotes the rectangular pulse that equals $1$ for $t \in [0, 1]$ and $0$ otherwise. Without loss of generality, we adopt constellations with unit average power, i.e., $\mathbb{E}\{|s_{n,m}|^2\} = 1, \forall n, \forall m$.

\newcounter{TempEqCnt}
\setcounter{TempEqCnt}{\value{equation}}
\setcounter{equation}{7}
\begin{figure*}[!t]
    \begin{equation}
        \tilde{y}_{m, q} [i] = \frac{\alpha_q}{\sqrt{N}} \sum_{n=0}^{N-1} s_{n,m-1} e^{\jmath \frac{2\pi}{N} ni} e^{\jmath 2\pi n \Delta_f (T_{\text{CP}}-\tau_q)}  e^{\jmath 2\pi (m-1) f_{\text{d},q} T_{\text{s}} } g_1(i) + \frac{\alpha_q}{\sqrt{N}} \sum_{n=0}^{N-1} s_{n,m} e^{\jmath \frac{2\pi}{N} n i} e^{-\jmath 2\pi n \Delta_f \tau_q}  e^{\jmath 2\pi m f_{\text{d},q} T_{\text{s}} } g_2(i).  \vspace{-3mm} \label{eq:echo_time}
    \end{equation}
    \begin{subequations} \label{eq:rx_symbol}
        \begin{align}
            y_{n,m}^{\text{free}} & = \sum_{q=1}^{Q} \tilde{\alpha}_q s_{n,m} e^{-\jmath 2\pi n \Delta_f \tau_q} e^{\jmath 2\pi m f_{\text{d},q} T_{\text{s}} }, ~~ y_{n,m}^{\text{ICI}}  = \!\!\sum_{q=\widetilde{Q}+1}^{Q} \frac{\alpha_q}{N} e^{\jmath 2\pi m f_{\text{d},q} T_{\text{s}} } \sum_{n'=0 \atop n' \neq n}^{N-1} s_{n',m} e^{-\jmath 2\pi n' \Delta_f \tau_q}\frac{ 1-e^{\jmath\frac{2\pi}{N}(n'-n)(l_q-N_{\text{CP}})} }{1-e^{\jmath\frac{2\pi}{N}(n'-n)}},  \!\! \\
            y_{n,m}^{\text{ISI}}  & = \!\!\!\sum_{q=\widetilde{Q}+1}^{Q}\!\!\! \frac{\alpha_q}{N} e^{\jmath 2\pi (m-1) f_{\text{d},q} T_{\text{s}} } \Big( \!(l_q \!-\! N_{\text{CP}}) s_{n,m-1} e^{\jmath 2\pi n \Delta_f (T_{\text{CP}}-\tau_q)} \!+\!\! \sum_{n'=0 \atop n' \neq n}^{N-1}\!\! s_{n',m-1} e^{\jmath 2\pi n' \Delta_f (T_{\text{CP}}-\tau_q)} \frac{ 1 \!-\! e^{\jmath\frac{2\pi}{N}(n'-n)(l_q-N_{\text{CP}})} }{1 \!-\! e^{\jmath\frac{2\pi}{N}(n'-n)}} \!\Big).
        \end{align}
    \end{subequations} \vspace{-2mm}
    \rule[-0pt]{18.2cm}{0.05em}
\end{figure*}
\setcounter{equation}{\value{TempEqCnt}}

\vspace{-1mm}
\subsection{Radar Echo Signal Model}
Next, we briefly describe the signal processing procedure at the sensing receiver. Unlike typical wireless communication scenarios, the sensing detection window begins immediately after signal transmission, ensuring that no nearby targets are missed. Suppose there exist $Q$ point targets at ranges $R_q$ with relative radial velocities $v_q$ and radar cross section (RCS) $\sigma_{\text{rcs}, q}$ for $q = 1, \dots, Q$. Then, the echo signal at the sensing receiver can be written as \vspace{-1mm}
\begin{equation}
    y(t) = \sum_{q=1}^{Q} \alpha_q x(t-\tau_q) e^{\jmath 2\pi f_{\text{d}, q} t} + z(t). \vspace{-1mm}  \label{eq:echo_sig}
\end{equation}
The amplitude $\alpha_q$, round-trip delay $\tau_q$, and Doppler shift $f_{\text{d}, q}$ of the $q$-th target are respectively given by \vspace{-1mm}
\begin{equation}
    \alpha_q = \sqrt{\frac{\sigma_{\text{rcs}, q} c_0^2 G_{\text{Tx}} G_{\text{Rx}}}{(4 \pi)^3 R_q^4 f_\text{c}^2}}, ~\tau_q = \frac{2 R_q}{c_0}, ~f_{\text{d}, q} = \frac{2 v_q f_{\text{c}}}{c_0}. \vspace{-1mm}
\end{equation}
In these equations, $c_0$ is the speed of light, $f_\text{c}$ represents the carrier frequency, and $G_{\text{Tx}}$, $G_{\text{Rx}}$ denote the transmit and receive antenna gains, respectively. The noise term $z(t) \sim \mathcal{CN}(0, \sigma^2)$ models the additive white Gaussian noise (AWGN), with noise power given by  $\sigma^2 = FkBT_{\text{temp}}$, where $F$ is the receiver's noise figure, $k$ is Boltzmann's constant, $B = N \Delta_f$ is the system bandwidth, and $T_{\text{temp}}$ is the equivalent noise temperature.

At the sensing receiver, the radar echo signal $y(t)$ is sampled at the sampling interval $T_{\text{sam}} = T/N$. The resulting discrete-time echo signal is expressed as \vspace{-1mm}
\begin{subequations}
    \begin{align}
        y [i] & \approx \frac{1}{\sqrt{N}} \sum_{q=1}^{Q} \alpha_q \sum_{m=0}^{M-1} \sum_{n=0}^{N-1} s_{n,m} e^{\jmath \frac{2\pi}{N} n (i - N_{\text{CP}} - mN_{\text{s}})}              \\
              & ~~~ \times e^{-\jmath 2\pi n \Delta_f \tau_q}  e^{\jmath 2\pi m f_{\text{d},q} T_{\text{s}} } g\Big(\frac{i \!-\! l_q \!-\! mN_{\text{s}}}{N_{\text{s}}}\Big) \!+\! z[i],
    \end{align}
\end{subequations}
where $l_q = [\tau_q / T_{\text{sam}}]$, $N_{\text{CP}} = T_{\text{CP}}/ T_{\text{sam}}$, $N_{\text{s}} = N + N_{\text{CP}}$, and $[\cdot]$ is the rounding operation. For the echo signal corresponding to the $q$-th target, the $m$-th symbol spans the sample indices from $i = m N_{\text{s}} + l_q$ to $i = (m+1) N_{\text{s}} + l_q -1$. After CP removal at the sensing receiver, the FFT demodulation is applied to the $N$ samples from $i = m N_{\text{s}} + N_{\text{CP}}$ to $i = (m+1) N_{\text{s}}$. Consequently, if the delay of the $q$-th target does not exceed the CP length, neither ISI nor ICI will be introduced. Under this condition, the echo signal from the $q$-th target can be expressed as \vspace{-1mm}
\begin{equation}
    \!\!y_{m,q} [i] = \!\sum_{n=0}^{N-1}\! \frac{\alpha_q s_{n,m}}{\sqrt{N}}  e^{\jmath \frac{2\pi}{N} n i}  e^{-\jmath 2\pi n \Delta_f \tau_q}  e^{\jmath 2\pi m f_{\text{d},q} T_{\text{s}} } g\Big(\frac{i}{N}\Big).\!\!  \vspace{-1mm}
\end{equation}
However, if the delay of the $q$-th target exceeds the CP length, i.e., $l_q > N_{\text{CP}}$, both ISI and ICI arise, leading to degraded sensing performance. As illustrated in Fig.~\ref{fig:echo_signal}, the echo signal corresponding to the $m$-th symbol contains residual contributions from the previous symbol, resulting in ISI. Furthermore, because the detection window for the $m$-th symbol does not span the entire symbol duration, subcarrier orthogonality is lost, thereby inducing ICI. In this case, the echo signal for the $q$-th target during the $m$-th symbol is given by (\ref{eq:echo_time}) at the top of next page, where $g_1(i) = g(\frac{i}{l_q - N_{\text{CP}}})$, $g_2(i) = g(\frac{i - l_q + N_{\text{CP}}}{N - l_q + N_{\text{CP}}})$.

To differentiate between ISI/ICI-free echoes and ISI/ICI-contaminated echoes, we partition the $Q$ target echoes into two sets based on their round-trip delays. Without loss of generality, we assume that the first $\widetilde{Q}$ targets have round-trip delays such that $l_q \leq N_{\text{CP}}$. Under this assumption, the echo signal at the sensing receiver can be expressed as \vspace{-2mm}
\begin{equation}
    y_m [i] = \sum_{q=1}^{\widetilde{Q}}  y_{m,q} [i] + \sum_{q=\widetilde{Q}+1}^{Q} \tilde{y}_{m, q} [i] + z_m [i], \vspace{-1mm}
\end{equation}
where the first summation corresponds to the ISI/ICI-free echoes from targets with $l_q \le N_{\text{CP}}$, and the second summation represents the ISI/ICI-contaminated echoes from targets with $l_q > N_{\text{CP}}$. After OFDM demodulation, the frequency-domain echo signal on the $n$-th subcarrier of the $m$-th symbol is given by \vspace{-1mm}
\begin{equation}
    y_{n,m} = y_{n,m}^{\text{free}} + \underbrace{y_{n,m}^{\text{ISI}} - y_{n,m}^{\text{ICI}} + z_{n,m}}_{y_{n,m}^{\text{IN}}} , \vspace{-1mm}
\end{equation}
where $y_{n,m}^{\text{free}}$, $y_{n,m}^{\text{ISI}}$, and $y_{n,m}^{\text{ICI}}$ denote the useful signal component, ISI, and ICI, respectively, and $y_{n,m}^{\text{IN}}$ represents the interference-plus-noise (IN) term. The explicit expressions of $y_{n,m}^{\text{free}}$, $y_{n,m}^{\text{ISI}}$, and $y_{n,m}^{\text{ICI}}$ are provided in (\ref{eq:rx_symbol}). To simplify notation and highlight the impact of CP length,  we define $\rho_q \triangleq (l_q-N_{\text{CP}})/N$ and introduce a generalized target coefficient as
\setcounter{equation}{9}
\begin{equation}
    \tilde{\alpha}_q \triangleq \begin{cases}
        \alpha_q           & ~~q = 1, \dots, \widetilde{Q};   \\
        (1-\rho_q)\alpha_q & ~~q = \widetilde{Q}+1, \dots, Q,
    \end{cases}
\end{equation}
where the attenuation factor $1-\rho_q$ accounts for the signal energy loss caused by the FFT window failing to fully capture the delayed echo when $l_q > N_{\text{CP}}$. According to the findings in \cite{Wang_TVT_2025}, \cite{Xu_arxiv_2025}, both ISI and ICI can be approximated as CSCG distributions when $N$ is large enough, i.e., $y_{n,m}^{\text{ISI}} \sim (0, P_{\text{ISI}})$, $y_{n,m}^{\text{ICI}} \sim (0, P_{\text{ICI}})$, where their variances are given by \cite{Wang_TVT_2025}
\begin{equation}
    P_{\text{ISI}} =  \sum_{q=\widetilde{Q}+1}^{Q} \rho_q |\alpha_q|^2, P_{\text{ICI}} = \sum_{q=\widetilde{Q}+1}^{Q} \rho_q (1-\rho_q)  |\alpha_q|^2.
\end{equation}

\vspace{-2mm}
\subsection{Parameter Estimation via RF and MF} \label{sec:RDM_RF_MF}
Many filtering strategies can be employed to estimate target parameters, among which the most prominent are the RF and MF methods. The RDMs obtained using MF- and RF-based methods can be expressed as
\begin{subequations} \label{eq:RDM_free_IN}
    \begin{align}
        \chi^{\text{RF}}(l, \nu) & = \sum_{q=1}^{Q} \mathcal{T}_q^{\text{RF}}(l, \nu) + \mathcal{L}^{\text{RF}}(l, \nu), \\
        \chi^{\text{MF}}(l, \nu) & = \sum_{q=1}^{Q} \mathcal{T}_q^{\text{MF}}(l, \nu) + \mathcal{L}^{\text{MF}}(l, \nu),
    \end{align}
\end{subequations}
where $\mathcal{T}_q^{\text{RF}}(l, \nu)$ and $\mathcal{L}^{\text{RF}}(l, \nu)$ represent the contributions of the $q$-th target and the IN term $y_{n,m}^{\text{IN}}$ under RF processing, respectively. Likewise, $\mathcal{T}_q^{\text{MF}}(l, \nu)$ and $\mathcal{L}^{\text{MF}}(l, \nu)$ denote the corresponding components under MF processing. To facilitate analysis, we assume that the delay $\tau_q$ and Doppler shift $f_{\text{d}, q}$ are integer multiples of the corresponding resolution, i.e., $\tau_q  = l_q /B$, $f_{\text{d},q} = \nu_q /T_{\text{obs}}$, where $T_{\text{obs}} = M T_{\text{s}}$ denotes the observation time. The spectral leakage due to fractional delay and Doppler can be mitigated by applying appropriate window functions \cite{Harris_Proce_1978}. Then, the explicit expressions of $\mathcal{T}_q^{\text{RF}}(l, \nu)$, $\mathcal{L}^{\text{RF}}(l, \nu)$, $\mathcal{T}_q^{\text{MF}}(l, \nu)$, and $\mathcal{L}^{\text{MF}}(l, \nu)$ can be written as
\begin{subequations}
    \begin{align}
        \mathcal{T}_q^{\text{RF}}(l, \nu) & =  \frac{\tilde{\alpha}_q}{\sqrt{MN}} \sum_{m=0}^{M-1} \sum_{n=0}^{N-1} e^{\jmath \phi_{n,m,q}}, \label{eq:Target_RDM_RF}                                             \\
        \mathcal{L}^{\text{RF}}(l, \nu)   & = \frac{1}{\sqrt{MN}} \sum_{m=0}^{M-1} \sum_{n=0}^{N-1} \frac{y_{n,m}^{\text{IN}}}{s_{n,m}}  e^{\jmath \frac{2\pi}{N} nl} e^{-\jmath \frac{2\pi}{M} m \nu},           \\
        \mathcal{T}_q^{\text{MF}}(l, \nu) & = \frac{\tilde{\alpha}_q}{\sqrt{MN}} \sum_{m=0}^{M-1} \sum_{n=0}^{N-1} |s_{n,m}|^2 e^{\jmath \phi_{n,m,q}},                                                           \\
        \mathcal{L}^{\text{MF}}(l, \nu)   & = \!\frac{1}{\sqrt{MN}} \!\!\!\sum_{m=0}^{M-1} \!\sum_{n=0}^{N-1}\! y_{n,m}^{\text{IN}} s_{n,m}^{\ast} e^{\jmath \frac{2\pi}{N} nl} e^{-\jmath \frac{2\pi}{M} m \nu},
    \end{align}
\end{subequations}
where we define $\phi_{n,m,q} \triangleq \frac{2\pi}{N} n(l-l_q) + \frac{2\pi}{M} m (\nu_q - \nu)$. In the following section, we analyze the statistical characteristics and performance of the RDM under both RF and MF processing.

\section{Sensing Performance Analysis}
\subsection{Statistical Characteristics of RDM under RF and MF} \label{sec:Statistic_RDM}
Assuming statistical independence among different targets, the second-order moment of the RDM under RF processing can be expressed as
\begin{equation}
    \!\!\!\!\mathbb{E}\{ |\chi^{\text{RF}}(l,\nu)|^2 \} \!=\! \!\sum_{q'=1}^{Q}\! \mathbb{E}\{ |\mathcal{T}_q^{\text{RF}}(l,\nu)|^2 \} + \mathbb{E}\{ |\mathcal{L}^{\text{RF}}(l,\nu)|^2 \}. \label{eq:RF_output1}
\end{equation}
From (\ref{eq:Target_RDM_RF}), it is evident that $\mathcal{T}_q^{\text{RF}}(l, \nu)$ is a deterministic function, whose squared magnitude can be written as
\begin{subequations}
    \begin{align}
        \!\!|\mathcal{T}_q^{\text{RF}}(l, \nu)|^2
         & = \frac{|\tilde{\alpha}_q|^2}{MN} \Big| \sum_{n=0}^{N-1}\!\! e^{\jmath \frac{2\pi}{N} n(l-l_q)} \!\!\sum_{m=0}^{M-1}\!\! e^{\jmath \frac{2\pi}{M} m (\nu_q - \nu)} \Big|^2 \\
         & = \frac{|\tilde{\alpha}_q|^2}{MN} |D_N(l-l_q)|^2 |D_M(\nu-\nu_q)|^2,
    \end{align}
\end{subequations}
where we define $D_N(x) \triangleq \frac{\sin(\pi x)}{\sin(\pi x/N)} e^{\jmath \pi (N-1)x/N}$. Regarding the contribution of the IN term, $\mathcal{L}^{\text{RF}}(l, \nu)$ can be approximated as a CSCG random variable by invoking the central limit theorem (CLT). With independent and identically distributed properties of ISI, ICI, and noise, we have
\begin{subequations}
    \begin{align}
          & \mathbb{E}\{ |\mathcal{L}^{\text{RF}}(l, \nu)|^2 \}  \notag                                                                                                                                                                   \\
        = & \frac{1}{MN} \sum_{m,m'} \sum_{n,n'} \frac{\mathbb{E}\{y_{n,m}^{\text{IN}} (y_{n',m'}^{\text{IN}})^{\ast}\}}{\mathbb{E}\{s_{n,m} s_{n',m'}^{\ast}\}}  e^{\jmath \frac{2\pi}{N} (n-n')l} e^{-\jmath \frac{2\pi}{M} (m-m') \nu} \\
        = & \frac{\sigma^2_{\text{IN}}}{MN} \sum_{m=0}^{M-1} \sum_{n=0}^{N-1} \mathbb{E}\{1/|s_{n,m}|^2\},                                                                                                                                \\
        = & \xi_{s} \sigma^2_{\text{IN}},
    \end{align}
\end{subequations}
where $\xi_s = \mathbb{E}\{1/|s_{n,m}|^2\}$, $\sigma^2_{\text{IN}} = P_{\text{ISI}}+P_{\text{ICI}}+\sigma^2$. Therefore, the second-order moment of $\chi^{\text{RF}}(l,\nu)$ is given by
\begin{equation} \label{eq:RF_output2}
    \!\!\!\mathbb{E}\{ |\chi^{\text{RF}}(l,\nu)|^2 \} \!=\! \begin{cases}
        MN |\tilde{\alpha}_q|^2 \!+\! \xi_s\sigma^2_{\text{IN}}, & (l, q) \!=\! (l_q, \nu_q); \\
        \xi_s\sigma^2_{\text{IN}}                                & \text{otherwise}.
    \end{cases}
\end{equation}

Similarly, the second-order moment of the RDM under MF processing $\chi^{\text{MF}}(l, \nu)$ can also be written in the form of (\ref{eq:RF_output1}). However, $\mathcal{T}_q^{\text{MF}}(l, \nu)$ is random due to the randomness of the modulated symbols $s_{n,m}$. Its second-order moment can be derived as
\begin{subequations}
    \begin{align}
          & \mathbb{E}\{ |\mathcal{T}_q^{\text{MF}}(l, \nu)|^2 \}   \notag                                                                                                            \\
        = & \frac{|\tilde{\alpha}_q|^2}{MN} \!\!\sum_{m,m'}\!\sum_{n,n'}\! \mathbb{E}\{|s_{n,m}|^2 |s_{n',m'}|^2\} e^{\jmath (\phi_{n,m,q}-\phi_{n',m',q})}                           \\
        = & \frac{|\tilde{\alpha}_q|^2}{MN} \sum_{m=0}^{M-1} \sum_{n=0}^{N-1} \mathbb{E}\{|s_{n,m}|^4 \} + \frac{|\tilde{\alpha}_q|^2}{MN} \sum_{m,m'\neq m} \sum_{n,n'\neq n} \notag \\
          & ~~~ \times \mathbb{E}\{|s_{n,m}|^2 |s_{n',m'}|^2\} e^{\jmath (\phi_{n,m,q}-\phi_{n',m',q})}                                                                               \\
        = & |\tilde{\alpha}_q|^2 \mu_4 + \frac{|\tilde{\alpha}_q|^2}{MN} \Big( \Big|\sum_{m,n} e^{\jmath \phi_{n,m,q}}\Big|^2 - MN\Big)                                               \\
        = & |\tilde{\alpha}_q|^2 (\mu_4\!-\!1) + \frac{|\tilde{\alpha}_q|^2}{MN} |D_N(l-l_q)|^2 |D_M(\nu-\nu_q)|^2,
    \end{align}
\end{subequations}
where $\mu_4 = \mathbb{E}\{|s_{n,m}|^4\}$ denotes the fourth-order moment of the symbols. Besides, $\mathcal{L}^{\text{MF}}(l, \nu)$ can also be approximated as a CSCG random variable, and the variance of $\mathcal{L}^{\text{MF}}(l, \nu)$ can be obtained by
\begin{subequations}
    \begin{align}
        \!\!\!\!\mathbb{E}\{ |\mathcal{L}^{\text{MF}}(l, \nu)|^2 \}
        = & \frac{1}{MN} \!\!\sum_{m=0}^{M-1} \sum_{n=0}^{N-1} \mathbb{E}\{ |y_{n,m}^{\text{IN}}|^2 \} \mathbb{E}\{|s_{n,m}|^2 \}\!\! \\
        = & \sigma^2_{\text{IN}}.
    \end{align}
\end{subequations}
Therefore, the second-order moment of $\chi^{\text{MF}}(l, \nu)$ can be expressed as
\begin{equation} \label{eq:MF_output}
    \!\mathbb{E}\{|\chi^{\text{MF}}(l, \nu)|^2\} \!=\! \begin{cases}
        (MN \!+\! \mu_4 \!-\! 1)|\tilde{\alpha}_q|^2 \!+\! \sigma^2_{\text{IN}}, \!\!\!\! & \!\!(l, q) \!=\! (l_q, \nu_q); \\
        (\mu_4 -1)|\tilde{\alpha}_q|^2 +\sigma^2_{\text{IN}} ,                 \!\! \!\!  & \!\!\text{otherwise}.
    \end{cases}
\end{equation}

\subsection{PSLR and ISLR of  RDM under RF and MF}
To evaluate the performance of the RDM, we adopt two key performance metrics: the PSLR and ISLR. The PSLR quantifies the ratio of the peak sidelobe level to the mainlobe level, whereas the ISLR measures the total sidelobe level relative to the mainlobe level \cite{Lu_APSARConf_2007}. For the $q$-th target, the PSLR and ISLR are respectively defined as
\begin{subequations}
    \begin{align}
        \gamma_q & \triangleq \frac{ \mathbb{E}\Big\{ \underset{(l,\nu)\in \mathcal{R}_{\text{s}}}{\max} |\chi(l,\nu)|^2 \Big\} }{\mathbb{E}\big\{ |\chi(l_q,\nu_q)|^2 \big\}}, \\
        \beta_q  & \triangleq \frac{ \mathbb{E}\big\{ \sum_{(l,\nu)\in \mathcal{R}_{\text{s}}} |\chi(l,\nu)|^2 \big\} }{\mathbb{E}\big\{ |\chi(l_q,\nu_q)|^2 \big\}},
    \end{align}
\end{subequations}
where $\mathcal{R}_{\text{s}} = \big\{ (l, \nu) \big| (l, \nu) \neq (l_q, \nu_q), \forall q \big\}$ is the sidelobe region.

Based on the statistical analysis provided in Sec.~\ref{sec:Statistic_RDM}, we now derive the analytical expressions of the PSLR and ISLR under both RF and MF processing. Specifically, over the sidelobe region $\mathcal{R}_{\text{s}}$, $|\chi^{\text{RF}}(l, \nu)|^2$ follows an exponential distribution with mean $\xi_s\sigma^2_{\text{IN}}$. Following \cite{Arnold_book_1992}, \cite{Eisenberg_Statist_2008}, the expectation of the peak sidelobe level under RF processing can be obtained by \vspace{-1mm}
\begin{equation}
    \mathbb{E}\Big\{ \underset{(l,\nu)\in \mathcal{R}_{\text{s}}}{\max} |\chi^{\text{RF}}(l,\nu)|^2 \Big\} = H_Q \xi_s \sigma^2_{\text{IN}}, \vspace{-1mm}
\end{equation}
where $H_Q = \sum_{q=1}^{MN-Q} 1/q$ denotes the harmonic number. Additionally, the expectation of the integrated sidelobe level can be expressed as \vspace{-1mm}
\begin{equation}
    \mathbb{E}\Big\{ \sum_{(l,\nu)\in \mathcal{R}_{\text{s}}} |\chi^{\text{RF}}(l,\nu)|^2 \Big\}
    = (MN-Q) \xi_s \sigma^2_{\text{IN}}. \vspace{-1mm}
\end{equation}
Combining the above results, the PSLR and ISLR of the RDM under RF processing are given by \vspace{-1mm}
\begin{subequations}
    \begin{align}
        \gamma_q^{\text{RF}} & = \frac{H_Q \xi_s \sigma^2_{\text{IN}}}{MN |\tilde{\alpha}_q|^2 + \xi_s \sigma^2_{\text{IN}}},    \\
        \beta_q^{\text{RF}}  & = \frac{(MN-Q) \xi_s \sigma^2_{\text{IN}}}{MN |\tilde{\alpha}_q|^2 + \xi_s \sigma^2_{\text{IN}}}.
    \end{align}
\end{subequations}

For MF processing, according to the result in (\ref{eq:MF_output}), the expectations of the peak and integrated sidelobe level under MF processing are given by \vspace{-2mm}
\begin{subequations}
    \begin{align}
        \!\!\mathbb{E}\Big\{ \!\underset{(l,\nu)\in \mathcal{R}_{\text{s}}}{\max}\! |\chi^{\text{MF}}\!(l,\nu)|^2 \!\Big\}
         & \!=\! H_Q \!\Big(\!(\mu_4\!-\!1) \!\sum_{q=1}^{Q} \!|\tilde{\alpha}_q|^2 \!+\! \sigma^2_{\text{IN}}\!\Big),\!\!\! \\
        \mathbb{E}\Big\{\!\! \sum_{(l,\nu)\in \mathcal{R}_{\text{s}}}\!\!\!\!\! |\chi^{\text{MF}}(l,\nu)|^2 \!\Big\}
         & \!=\! (MN\!-\!Q) \Big((\mu_4\!-\!1) \sum_{q=1}^{Q} |\tilde{\alpha}_q|^2 \!+\! \sigma^2_{\text{IN}}\Big).
    \end{align}
\end{subequations}
Thus, the PSLR and ISLR of the RDM under MF processing can be written as \vspace{-1mm}
\begin{subequations}
    \begin{align}
        \gamma_q^{\text{MF}} & = \frac{H_Q \Big((\mu_4-1) \sum_{q=1}^{Q} |\tilde{\alpha}_q|^2 + \sigma^2_{\text{IN}}\Big)}{MN |\tilde{\alpha}_q|^2 + (\mu_4-1) \sum_{q=1}^{Q} |\tilde{\alpha}_q|^2  + \sigma^2_{\text{IN}}},    \\
        \beta_q^{\text{MF}}  & = \frac{(MN-Q) \Big((\mu_4-1) \sum_{q=1}^{Q} |\tilde{\alpha}_q|^2 + \sigma^2_{\text{IN}}\Big)}{MN |\tilde{\alpha}_q|^2 + (\mu_4-1) \sum_{q=1}^{Q} |\tilde{\alpha}_q|^2  + \sigma^2_{\text{IN}}}.
    \end{align}
\end{subequations}

\subsection{Sensing Performance Comparison Between RF and MF}
\begin{table}[!t]
    \centering
    \footnotesize
    \caption{$\xi_s$ and $\mu_4$ values of typical constellations}
    \label{Tab:table_xi}
    \begin{tabular}{ccccc}
        \toprule
        Constellation & PSK & 16-QAM & 256-QAM & 1024-QAM \\
        \midrule
        $\xi_s$       & 1   & 1.8889 & 3.4374  & 4.1673   \\
        $\mu_4$       & 1   & 1.3199 & 1.3953  & 1.3989   \\
        \bottomrule
    \end{tabular}
\end{table}

As shown in Table \ref{Tab:table_xi}, for constant-modulus modulations such as PSK, both $\xi_s$ and the fourth-order moment $\mu_4$ are equal to one. This condition yields identical PSLR and ISLR expressions for the RDM under both RF and MF processing, indicating that the two methods are theoretically equivalent in sensing performance for this scenario. However, with a non-constant-modulus modulation such as quadrature amplitude modulation (QAM), the values of $\xi_s$ and $\mu_4$ exceed one, leading to notable performance differences between the RF and MF approaches. The RF-based method effectively suppresses inter-target interference (ITI), which is advantageous in scenarios with closely spaced targets or large disparities in target RCS. Nevertheless, the increased $\xi_s$ associated with high-order QAM produces a proportional amplification of the IN power, thereby degrading the sensing performance. This degradation is further exacerbated when the CP is insufficient, since ISI and ICI elevate the interference floor. In contrast, the MF-based method does not amplify the IN power but instead introduces ITI. Although $\mu_4$ grows more slowly than $\xi_s$, the ITI induced by MF can still be detrimental when weak targets coexist with strong ones, potentially obscuring the weaker targets. Thus, while RF and MF techniques are equivalent under PSK modulation, their trade-offs diverge under QAM: the RF approach achieves better ITI suppression at the cost of an elevated noise floor, whereas the MF approach maintains a lower noise floor but suffers from stronger ITI.

\section{Simulation Results}
\begin{table}[!t]
    \footnotesize
    \centering
    \caption{Simulation Parameters} \label{Tab:parameters}
    \begin{tabular}{ccc}
        \toprule
        \textbf{Parameter}    & \textbf{Symbol}                  & \textbf{Value} \\
        \midrule
        Carrier frequency     & $f_{\text{c}}$                   & 28 GHz         \\
        Subcarrier spacing    & $\Delta_f$                       & 120 kHz        \\
        Number of subcarriers & $N$                              & 256            \\
        Number of symbols     & $M$                              & 128            \\
        Symbol duration       & $T$                              & 8.33 $\mu$s    \\
        Normal CP length      & $T_{\text{CP}}$                  & 0.59 $\mu$s    \\
        Antenna gain          & $G_{\text{Tx}}$, $G_{\text{Rx}}$ & 25.8 dBi       \\
        Noise figure          & $F$                              & 3 dB           \\
        Reference temperature & $T_{\text{temp}}$                & 290 K          \\
        \bottomrule
    \end{tabular} \vspace{-0.2 cm}
\end{table}

\begin{figure}[!t]
    \vspace{-2mm}
    \centering
    \includegraphics[width = 2.8 in]{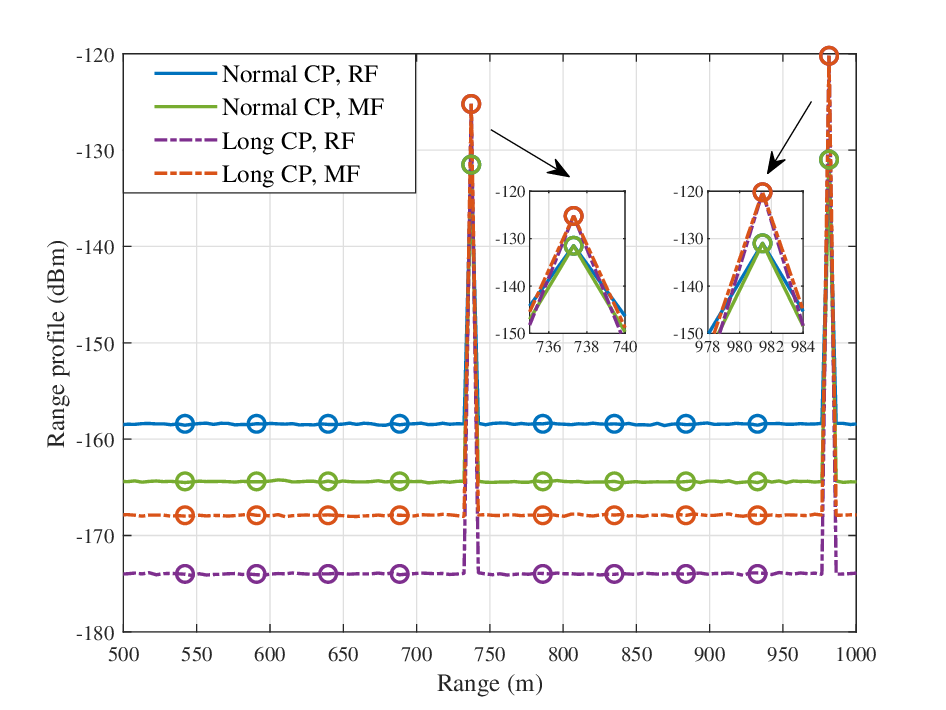}
    \vspace{-1mm}
    \caption{The range profile under RF and MF processing.}
    \label{fig:range_profile} \vspace{-0.2 cm}
\end{figure}

\begin{figure}[!t]
    \vspace{-2mm}
    \subfigcapskip = -4pt
    \centering
    \subfigure[PSLR.]{
        \includegraphics[width = 2.8 in]{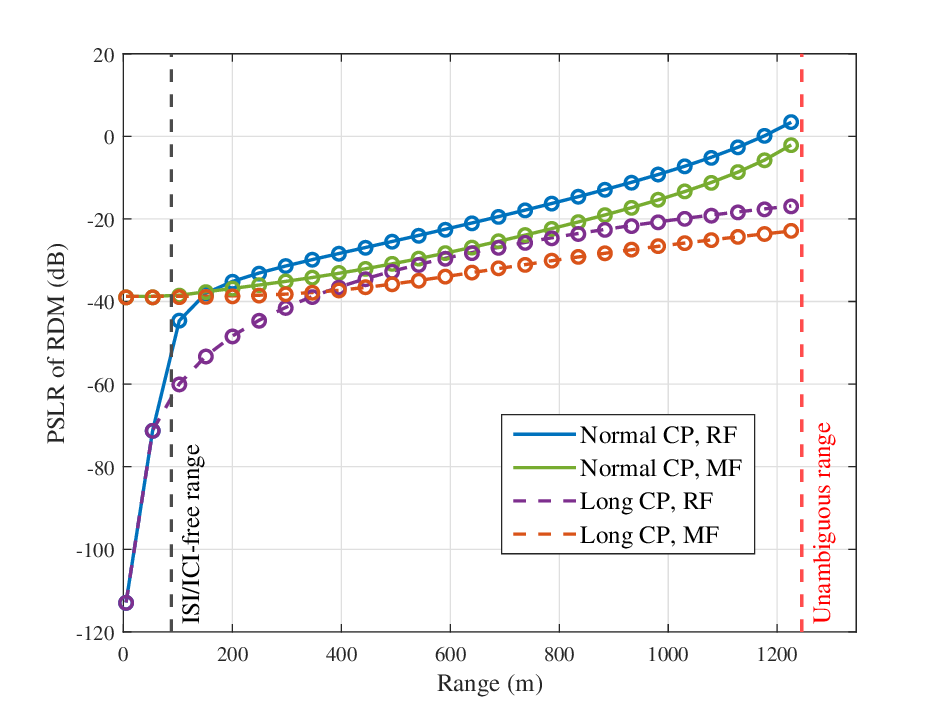}
    }
    \subfigure[ISLR.]{
        \includegraphics[width = 2.8 in]{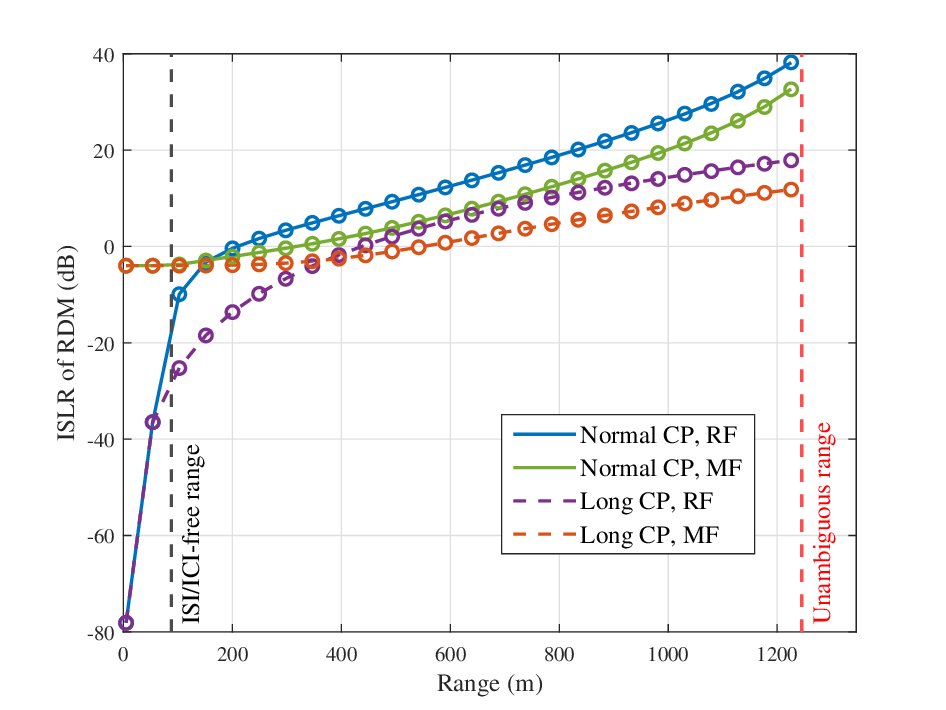}
    }
    \vspace{-0.1 cm}
    \caption{PSLR and ISLR of the RDM versus target range under RF and MF processing.}\label{fig:PSLR_ISLR} \vspace{-0.3 cm}
\end{figure}

In this section, we present simulation results to validate our theoretical analysis. The simulation parameters are provided in Table \ref{Tab:parameters}. Throughout the simulations, 1024-QAM is employed. Moreover, all simulation results are attained by averaging over 5000 random realizations.

Fig.~\ref{fig:range_profile} shows the range profiles for two targets processed by the RF and MF methods under two CP configurations: a normal CP and a long CP. Here, the long CP length is set equal to the full symbol duration $T$. As evident from Fig.~\ref{fig:range_profile}, the simulation results (plotted as lines) closely match the theoretical values (depicted by circular markers), validating the accuracy of our analytical derivations in Sec.~\ref{sec:Statistic_RDM}. From our calculations, the maximum ISI/ICI-free range is 87.9m for the normal CP configuration and 1250m for the long CP configuration. Because both targets (at 732.4m and 976.5m) lie well beyond 87.9m, the normal CP case experiences severe ISI/ICI: the target mainlobes are notably attenuated and the sidelobes are significantly elevated. Moreover, in the normal CP scenario, the RF processing further amplifies the IN power (due to a larger $\xi_s$), resulting in even higher sidelobe levels compared to the MF. In contrast, with the long CP, the targets fall well within the 1250m interference-free range, so no ISI/ICI distortion is observed. Under these conditions, the RF method achieves a lower sidelobe level than the MF, owing to its superior suppression of ITI.

Next, Fig.~\ref{fig:PSLR_ISLR} illustrates the PSLR and ISLR versus target range for RF and MF processing, with lines representing simulation results and circular markers denoting theoretical values. The perfect match validates our analytical derivations. As expected, PSLR and ISLR gradually increase with the target range under both CP configurations. Notably, once the target range surpasses the ISI/ICI-free limit of normal CP, both PSLR and ISLR rise sharply compared to the long CP case, with degradation accelerating at longer distances due to accumulated ISI/ICI. In addition, for short target ranges (high echo SNR), ITI is the dominant factor limiting the estimation performance. In this regime, the randomness of the modulated symbols causes MF processing to introduce additional target leakage, whereas RF processing effectively avoids this issue. Consequently, RF yields lower PSLR and ISLR than MF, indicating superior performance at short ranges. Conversely, at longer ranges with weaker echoes, RF's inherent noise amplification ($\xi_s > 1$) significantly elevates the noise floor, causing its PSLR and ISLR to exceed those of MF and thus degrading RF performance.

\begin{figure}[!t]
    \vspace{-2mm}
    \subfigcapskip = -4pt
    \centering
    \subfigure[Range estimation RMSE.]{
        \includegraphics[width = 2.8 in]{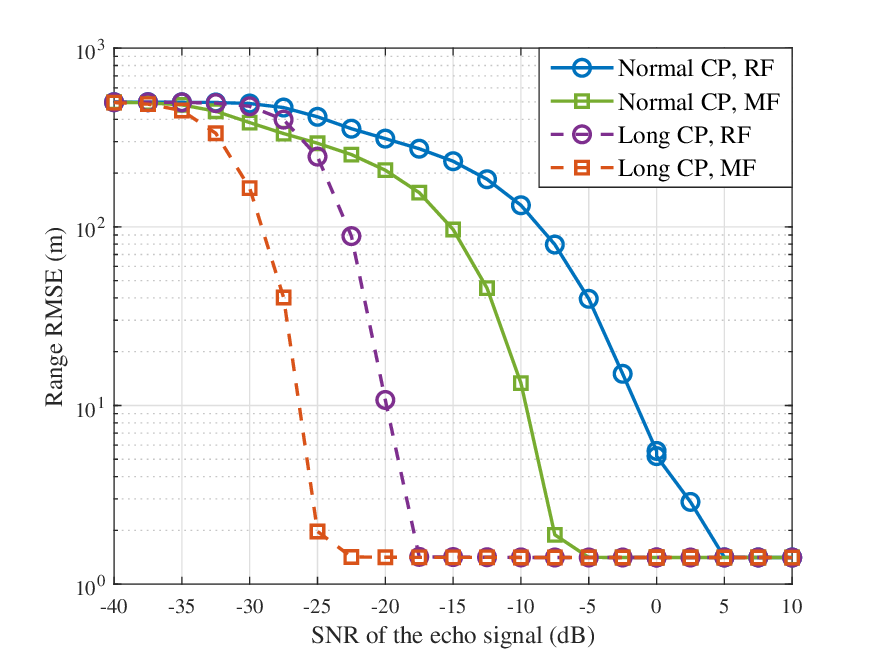}
    }
    \subfigure[Velocity estimation RMSE.]{
        \includegraphics[width = 2.8 in]{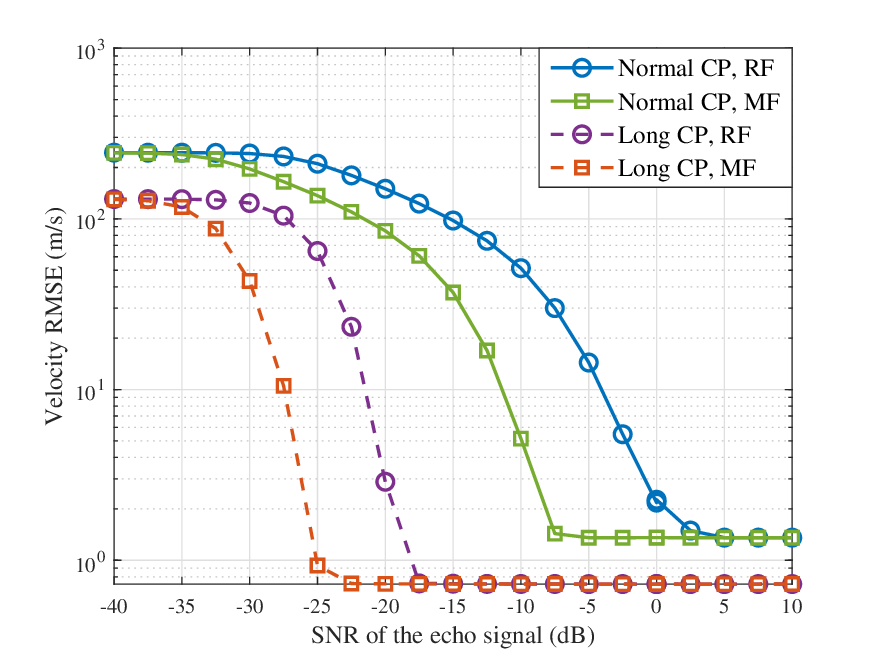}
    }
    \vspace{-0.1 cm}
    \centering
    \caption{RMSE for range and velocity estimation versus the sensing SNR.}\label{fig:RMSE_SNR} \vspace{-0.4 cm}
\end{figure}
Finally, Fig.~\ref{fig:RMSE_SNR} quantitatively evaluates the parameter estimation performance by plotting the root-mean-square-error (RMSE) of range and velocity versus the echo SNR under RF and MF processing. The target is assumed to be uniformly distributed within the unambiguous range and Doppler, with a fixed RCS of 5dBsm. As expected, in this single-target scenario, where no ITI is present, the MF approach achieves a lower RMSE than the RF approach for both range and velocity estimation. Using a normal CP results in significantly higher RMSE values than a long CP for both range and velocity, indicating severe performance degradation due to ISI and ICI when the CP duration is insufficient. Furthermore, the extended observation time provided by a long CP yields a finer Doppler resolution, further reducing the velocity estimation RMSE compared to the normal CP case.

\section{Conclusion}
This work has quantified the impact of CP length on the sensing performance by developing a comprehensive model that captured insufficient-CP-induced ISI and ICI in multi-target scenarios. Our unified analysis provided closed-form expressions for the second-order statistical characterizations of the RDM under both RF and MF processing. The derived expressions for PSLR and ISLR revealed how CP length, modulation order, and filter choice jointly determined the trade-off between noise amplification and ITI. Simulation results validated the theoretical analysis and demonstrated that, while RF outperformed MF in suppressing ITI for near-range targets, its noise amplification effect causes MF to be preferable for long range sensing. These insights provided practical guidelines for selecting filtering strategies and modulation schemes in OFDM-ISAC systems to optimize the sensing performance. This initial work motivates further investigation of advanced signal processing techniques to mitigate the ISI and ICI caused by insufficient CP, thereby enhancing the robustness and accuracy of parameter estimation in OFDM-ISAC systems.

\end{document}